\documentclass[aps,reprint]{revtex4-2}
\usepackage{amsmath}
\usepackage{amssymb}
\usepackage{graphicx}
\usepackage{dcolumn}
\usepackage{hyperref}
\usepackage{bm}
\usepackage{tikz}
\usetikzlibrary{arrows}
\begin{document}
\title{\textbf{Arbitrarily precise arrival time measurements in quantum mechanics} } 
\author{Lawrence Frolov}
\affiliation{Department of Mathematics, Rutgers University (New Brunswick)}
\date{\today}
\begin{abstract}
    The quantum Zeno effect is often regarded as an obstruction to precise arrival time measurements in quantum mechanics. Here, an arbitrarily precise arrival time measurement procedure is constructed using a localized detection process and a suitably chosen boundary condition. The arrival time of an incoming particle is recorded in the position of a clock particle that is emitted by the apparatus upon detection. A non-zero probability of arrival is shown to survive even in the limit as exact arrival time measurement precision is achieved. In this limit, it is also found that the interaction between the incoming particle and the detector is described by an absorbing boundary condition. This justifies the claim of Tumulka that absorbing boundary conditions may be used to model idealized detectors capable of registering particles at the instant of their arrival.
\end{abstract}
\maketitle
\section{Introduction}
Suppose that a single particle wave function $\psi_0$ is prepared in $x<0$ with a detector waiting to measure the time of arrival at $x=0$. Assuming that their interaction is negligible outside of $x \geq 0$, the dynamics of the incoming particle is governed by a Schr\"odinger equation that is decoupled from the detector in $x<0$. In contrast, a dominant interaction must take place in $x \geq 0$ that rapidly transitions the apparatus from its ``ready" state to a ``post-detection" state residing in one of many channels $\mathcal{H}_t$ that are labeled by the recorded arrival time.
\par
Arrival time measurements are notoriously difficult to treat in quantum mechanics and have remained the subject of active debate for decades. One particular source of contention surrounds a counterintuitive prediction of many theoretical models for the arrival time measurement procedure: the probability of the particle's arrival at $x=0$ necessarily vanishes in the limit as exact arrival time measurement precision is achieved \cite{ALLCOCK,Friedman,watchdog, Aharonov, MugaBook,Yearsley, IdealizedClocks}. The dominant interaction in $x\geq0$ has the unintended effect of reflecting a fraction of the incoming wave, and this effect intensifies so that all wave functions are completely reflected as the arrival time measurement procedure is made arbitrarily precise. This phenomenon can be regarded as a particular instance of the quantum Zeno effect \cite{Zeno, FacchiChap}, and it has led to much speculation about a fundamental principle in quantum mechanics that denies the possibility of precise ``continuous observation". 
\par
The purpose of this article is to show, using a simple detector model, that quantum mechanics can accommodate an arbitrarily precise arrival time measurement procedure without the complete reflection of all incoming wave functions. The approach taken here differs from others who proposed normalizing the vanishing probability distribution \cite{DisclosingZeno, DisclosingCompl} or using quantum stroboscopy \cite{Stroboscopy} to obtain a non-zero \textit{conditional} probability distribution for the arrival time. A family of arrival time measurement procedures is constructed that admits a non-zero probability for the registered arrival, even in the limit as exact precision is achieved.
\par
A description of the procedure is as follows. The detector employs a clock particle of mass $M$ to record the arrival time and contains a two-level system with energy difference $\hbar \omega$ that transitions to the lower energy state upon detection. In contrast to previously studied ``quantum clock" models \cite{Peres,Aharonov,Clockreview, IdealizedClocks}, here the incoming particle is \textit{absorbed} upon detection, and, at that same instant, the clock particle is \textit{emitted}. The uncertainty in the clock configuration is made arbitrarily small at the instant of emission through a suitable localization of the absorption-emission process. This localization is enforced by placing a wall near $x=0$, modeled by a Neumann boundary condition \cite{Meaning}, so that the incoming particle cannot penetrate too far into the detector region.   
\par
A family of exact arrival time measurement procedures is shown to emerge as the detection process is completely localized at $x=0$ and as $\omega,M\to \infty$ with $2\hbar\omega/M\to v_c^2>0$. In this limit, arrivals at $x=0$ are registered instantly by the detector and the emitted clock particle is classically transported with constant velocity $v_c$ so that its position permanently records the arrival time. The localized interaction between the incoming particle and the apparatus is also shown to converge to an absorbing boundary condition (ABC) in this limit. ABCs have been studied as localized detector models by several authors \cite{Werner1987,Parameter,thesis}, and were notably proposed by Tumulka \cite{TUMULKA} as describing idealized detectors capable of registering a particle at the instant of arrival. That proposal is verified here. These localized interactions do generate a partial reflection of every incoming wave function, supporting the conjecture of Allcock \cite{ALLCOCK} and Aharonov et al. \cite{Aharonov} that exceptionally precise arrival time measurements produce a \textit{partial} quantum Zeno effect that distorts the arrival outcomes in an apparatus-dependent manner.
\par
This paper is organized as follows: in Section \ref{main} a family of arbitrarily precise arrival time measurement procedures is constructed, and in Section \ref{disc} an analysis of their partial quantum Zeno effect is provided.
\section{Arbitrarily precise arrival time measurement procedures}\label{main}
A direct measurement of the arrival time at $x=0$ consists of three distinct steps: the registration of the particle exiting $x<0$; the formation of a macroscopic record of the arrival time; and the preservation of this record for future observation. Achieving exact precision requires that the following conditions hold for any incoming wave function:
\begin{enumerate}
    \item[(C1)] the registration process in $x \geq 0$ must be sufficiently rapid, preferably instantaneous, so that parts of the wave function exiting $x< 0$ are quickly transferred to the post-detection state space;
    \item[(C2)] parts of the wave function exiting $x<0$ must be brought to the correct time-labeled channel $\mathcal{H}_t$;
    \item[(C3)] records must remain stable. States residing in $\mathcal{H}_t$ may not transition to any state residing in a different time-labeled channel $\mathcal{H}_{t'}$, or back to a ready state, otherwise the registration process may be repeated. Detections must be effectively irreversible.
\end{enumerate}
A family of measurement procedures that satisfy the above conditions shall now be constructed. Motivated by the irreversible detector model of Halliwell \cite{Halliwell}, the detector here contains a two-level system with states $|\text{R}\rangle$ and $|\text{PD}\rangle$ representing the ready state and the post-detection state, respectively. The Hamiltonian of the detector is 
\begin{eqnarray}
    \hat{H}_D=0 |\text{R}\rangle \langle \text{R}| -\hbar \omega |\text{PD}\rangle \langle\text{PD}|,
\end{eqnarray}
so that $|\text{R}\rangle $ and $|\text{PD} \rangle$ are energy eigenstates of $\hat{H}_D$, with eigenvalues $0$ and $-\hbar\omega$, respectively. Halliwell achieved irreversibility by coupling the system to a large number of harmonic oscillators, but here it shall follow directly from the stability of the arrival time recording device. 
\par
The particle-detector state is initially prepared as a pure-product $\Psi_0=\psi_0\otimes |\text{R}\rangle$, where $\psi_0$ is supported in $x<0$. $\Psi$ evolves unitarily in a Hilbert space $\mathcal{H}_\text{ND}\oplus\mathcal{H}_\text{PD}$ that contains two sectors. The dynamics in the ready state space $\mathcal{H}_\text{ND}=L^2((-\infty,\varepsilon))\otimes |\text{R}\rangle$ model an incoming particle of mass $m$ while the detector remains ready. Detections are allowed to occur anywhere within the interval $(0,\varepsilon)$, but will eventually become concentrated at $x=0$ after taking $\varepsilon \to 0$. A wall, modeled through a Neumann boundary condition, is placed at $x=\varepsilon$ to ensure that the incoming particle does not penetrate past that point. The dynamics in the post-detection state space $\mathcal{H}_\text{PD}=L^2((-\infty,\varepsilon))\otimes |\text{PD}\rangle$ model an emitted clock particle of mass $M$, so the particle Hamiltonian is
\begin{equation}
    \hat{H}_p=-\frac{\hbar^2}{2m}\partial_x^2~|\text{R}\rangle\langle \text{R}| - \frac{\hbar^2}{2M}\partial_x^2~|\text{PD}\rangle \langle\text{PD}|.
\end{equation}
The absorption of the incoming particle and subsequent emission of the clock particle is implemented through an interaction Hamiltonian that couples the two sectors
\begin{equation}
    \hat{H}_I= W(x)|\text{R}\rangle \langle \text{PD}|+W^*(x) |\text{PD} \rangle\langle\text{R}|.
\end{equation}
Here, $W(x)$ is a complex-valued function supported in the detection region $(0,\varepsilon)$. Expanding the total wave function as $\Psi_t=\psi_t\otimes |\text{R}\rangle + \phi_t\otimes |\text{PD}\rangle$, one finds that $i\hbar \partial_t\Psi=(\hat{H}_D+\hat{H}_p+\hat{H}_I)\Psi$ can be written as a coupled system of one-body Schr\"odinger equations
\begin{equation}
    i\hbar \partial_t\begin{bmatrix}
        \psi\\ \phi
    \end{bmatrix}=\begin{pmatrix}
        -\frac{\hbar^2}{2m}\partial_x^2 &W(x)
        \\
        W^*(x)&-\frac{\hbar^2}{2M}\partial_x^2-\hbar\omega
    \end{pmatrix}\begin{bmatrix}
        \psi
        \\
        \phi
    \end{bmatrix}.
\end{equation}
These evolution equations are supplemented with the Neumann boundary conditions $\partial_x\psi(\varepsilon)=0=\partial_x\phi(\varepsilon)$.
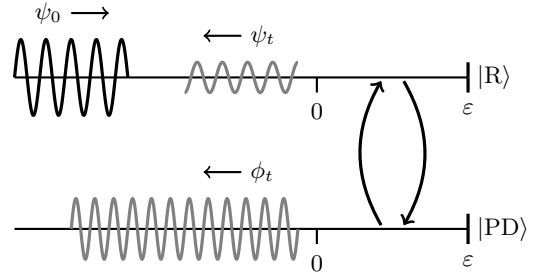
\begin{figure}[h]
\centering
\begin{tikzpicture}
  \draw[-][thick] (0, 0) -- (6, 0) node[right] {$|\text{PD}\rangle$};
  \draw[-][thick] (0, 2) -- (6, 2) node[right] {$|\text{R}\rangle$};
  \draw[->][very thick] (4.85, 0.05) to [bend left=30](4.85, 1.95) ;
  \draw[->][very thick] (5.15, 1.95) to [bend left=30](5.15, 0.05) ;
  \draw[very thick,variable=\t,domain=0:1.5,samples=1000]
    plot (\t,{0.5*sin(3*\t*360)+2},0);
    \draw[very thick,gray, variable=\t,domain=2.25:3.75,samples=1000]
    plot (\t,{0.2*sin(3*\t*360)+2},0);
    \draw[very thick,gray, variable=\t,domain=0.75:3.75,samples=2000]
    plot (\t,{0.4*sin(4*\t*360)},0);
    \draw[->][thick] (0.75, 2.85)node[left] {$\psi_0$} -- (1.25, 2.85);
    \draw[<-][thick] (2.5, 2.55) -- (3, 2.55) node[right] {$\psi_t$};
    \draw[<-][thick] (2.5, 0.75) -- (3, 0.75) node[right] {$\phi_t$};
    
    \draw [thick] (4, 0) -- ++(0, -.2) ++(0, -.15) node [below, outer sep=0pt, inner sep=0pt] {\small\(0\)};
    \draw [very thick] (6, 0) -- ++(0, -.2) ++(0, -.15) node [below, outer sep=0pt, inner sep=0pt] {$\varepsilon$};
    \draw [very thick] (6, 0) -- ++(0, .2) ++(0, -.15) node [below, outer sep=0pt, inner sep=0pt] {};
    \draw [thick] (4, 2) -- ++(0, -.2) ++(0, -.15) node [below, outer sep=0pt, inner sep=0pt] {\small\(0\)};
    \draw [very thick] (6, 2) -- ++(0, -.2) ++(0, -.15) node [below, outer sep=0pt, inner sep=0pt] {$\varepsilon$};
    \draw [very thick] (6, 2) -- ++(0, .2) ++(0, -.15) node [below, outer sep=0pt, inner sep=0pt] {};
  
\end{tikzpicture}
\caption{A part of the incoming wave is transferred to the post-detection state space, but some parts are reflected instead. }
\label{Figure}
\end{figure}
\par
The amplitude of $W(x)$ regulates the rate at which parts of the incoming wave function entering $(0,\epsilon)$ are transferred to the post-detection state space. However, it has been known since the work of Allcock \cite{ALLCOCK} and Kraus \cite{watchdog} that trying to achieve an infinitely rapid registration process by taking $|W(x)| \to \infty$ in $(0,\varepsilon)$ leads to the quantum Zeno effect. The ready state $\psi$ satisfies the Dirichlet boundary condition $\psi(0)=0$ in this limit, causing all incoming wave functions to be completely reflected at $x=0$ with zero probability of detection.
\par
This shall be avoided here by localizing the registration process at $x=0$ in the limit as it is made infinitely rapid. This is accomplished by setting $W(x)=W\theta(x)/\varepsilon$ for $W \in \mathbb{C}$ and taking $\varepsilon \to 0$. Of course, localizing the registration process in this manner may allow sufficiently fast wave packets to quickly enter and exit the detection region without giving sufficient time for an interaction to occur. This concern will be addressed later by demonstrating that $|W|$, which controls the registration rate relative to the size of the detection region, can be taken to infinity along with another apparatus parameter while still retaining a non-zero probability of detection.
\par
After repeating similar steps to those taken in \cite{Albeverio, ABCCOMPLEX}, one finds that the limiting dynamics as $\varepsilon \to 0$ is governed by a system of free Schr\"odinger equations in $x<0$ coupled through their boundary conditions at $x=0$
\begin{subequations}
\begin{equation}
    i\hbar \partial_t\begin{bmatrix}
        \psi\\ \phi
    \end{bmatrix}=\begin{pmatrix}
        -\frac{\hbar^2}{2m}\partial_x^2 &0
        \\
        0&-\frac{\hbar^2}{2M}\partial_x^2-\hbar\omega
    \end{pmatrix}\begin{bmatrix}
        \psi
        \\
        \phi
    \end{bmatrix},
\end{equation}
    \begin{eqnarray}
\partial_x\psi(0)=-\frac{2m}{\hbar^2}W\phi(0), ~~ \partial_x\phi(0)=-\frac{2M}{\hbar^2}W^*\psi(0).
    \end{eqnarray}
\end{subequations}
As is, this arrival time measurement procedure still admits several defects. For a particle to act as a reliable clock, it must travel with a reliable velocity. However, the velocity, loosely speaking, of the emitted particle depends on the momentum of the incoming particle according to $v_\text{em}=p_\text{em}/M=\sqrt{(p^2_\text{in}/m + 2\hbar\omega)/M}$. Moreover, the position of the emitted particle does not act as a stable arrival time record because the post-detection dynamics allows parts of $\phi$ to disperse and even propagate back to the non-detection state space.
\par
These defects can be cured by taking the limit as $\omega,M\to \infty$ with $\hbar \omega=Mv_c^2/2$ for some fixed velocity $v_c>0$. Given that  $\psi\big{|}_{t=0}=\psi_0$ and $\phi\big{|}_{t=0}=0$, it is shown in the appendix that the limiting dynamics converge and are governed by
\begin{subequations}\label{stopwatch}
\begin{equation}
    i\hbar \partial_t\begin{bmatrix}
        \psi\\ \phi
    \end{bmatrix}=\begin{pmatrix}
        -\frac{\hbar^2}{2m}\partial_x^2 &0
        \\
        0&i\hbar v_c\partial_x-2\hbar \omega 
    \end{pmatrix}\begin{bmatrix}
        \psi
        \\
        \phi
    \end{bmatrix},
\end{equation}
    \begin{eqnarray}
\partial_x\psi(0)=i\kappa\psi(0), \quad \phi(0)&=-i\frac{2W^*}{\hbar v_c}\psi(0),
    \end{eqnarray}
\end{subequations}
where $\kappa=4m|W|^2/(v_c\hbar^3)>0$. The detection process is now effectively irreversible, and its singular interaction with the incoming particle is encapsulated by the absorbing boundary condition $\partial_x\psi(0)=i\kappa\psi(0)$. This boundary condition ensures that probability irreversibly flows out of the ready state space through $x=0$ with 
\begin{eqnarray}
    -\frac{d}{dt}||\psi||_{L^2}^2&&=j_x(0)=\frac{\hbar}{m}\text{Im}(\psi^*(0)\partial_x\psi(0)) \nonumber
    \\
    &&=\frac{\hbar \kappa}{m}\left |\psi(0)\right|^2\geq0. 
\end{eqnarray}
Parts of $\psi$ that exit $x<0$ are instantly transferred to the post-detection state space, and there, they undergo a classical transport to the left with constant velocity $v_c$
\begin{equation}
    \phi_t=\begin{cases}
        \frac{-2iW^*}{\hbar v_c}e^{-2i\omega x/v_c}\psi_{t+x/v_c}(0) \quad &\text{for }x>-v_ct
        \\
        0 \quad &\text{for }x<-v_ct
    \end{cases}
\end{equation}
The position of the emitted clock particle precisely encodes the time of arrival, in that the distance of the emitted particle from $x=0$ divided by $v_c$ records the time that has \textit{passed} since the incoming particle arrived at $x=0$. This record remains stable for all time because the transport does not allow interference between post-detection states with disjoint support. Moreover, for any time $t>0$, the Born rule for the probability that the clock is ``on the record" for an arrival having occurred at $\tau\in(t_1,t_2)\subset (0,t)$ agrees with the flux distribution
\begin{align}
    \mathbb{P}(\tau&\in (t_1,t_2))=\mathbb{P}(x\in (v_c(t_1-t),v_c(t_2-t)) \nonumber
    \\
    &=\int_{v_c(t_1-t)}^{v_c(t_2-t)}dx ~|\phi_t(x)|^2 \nonumber
    \\
    &=\int_{t_1}^{t_2}d\tau ~\frac{\hbar \kappa}{m}|\psi_\tau(0)|^2=\int_{t_1}^{t_2}d\tau ~j_x(0).
\end{align}
This arrival time probability distribution is known to be given by a positive operator-valued measure \cite{Werner1987,Existence} and admits an energy-time uncertainty relation \cite{Kiukas}, in that the variance of the arrival time distribution multiplied by the variance in the initial energy distribution of the incoming particle is bounded from below. 
\par
The infinitely oscillating term appearing in the solution formula for $\phi_t$ may raise eyebrows. This term does not influence the statistics of recorded arrival outcomes and is a consequence of the obscene amount of energy that the two-level system dumps onto the emitted particle. It also ensures that the probability current for the post-detection state converges with 
\begin{equation*}
    \frac{\hbar}{M} \text{Im}\left(\phi^* \partial_x\phi \right)\xrightarrow[M=2\hbar \omega/v_c^2]{\omega \to \infty} -v_c|\phi|^2.
\end{equation*}

\section{The partial quantum Zeno effect}\label{disc}
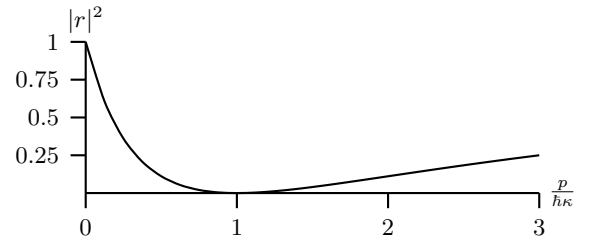
\begin{figure}[h]
\centering
\begin{tikzpicture}
  \draw[-][thick] (0, 0) -- (6, 0) node[right] {$\frac{p}{\hbar \kappa}$};
  \draw[-][thick] (0, 0) -- (0, 2) node[above] {$|r|^2$};
    \foreach \x in {0, ..., 3} {
        \draw [thick] (2*\x, 0) -- ++(0, -.2) ++(0, -.15) node [below, outer sep=0pt, inner sep=0pt] {\small\(\x\)};
    }
  \draw [thick] (0, 1/2) -- ++(-.2, 0) ++(-.15, 0) node [left, outer sep=0pt, inner sep=0pt] {\small\(0.25\)};
  \draw [thick] (0, 1) -- ++(-.2, 0) ++(-.15, 0) node [left, outer sep=0pt, inner sep=0pt] {\small\(0.5\)};
  \draw [thick] (0, 3/2) -- ++(-.2, 0) ++(-.15, 0) node [left, outer sep=0pt, inner sep=0pt] {\small\(0.75\)};
  \draw [thick] (0, 2) -- ++(-.2, 0) ++(-.15, 0) node [left, outer sep=0pt, inner sep=0pt] {\small\(1\)};
  
  \draw[scale=2, domain=0:3, smooth, variable=\x, black][thick] plot ({\x}, {((\x-1)/(\x+1))^2});
\end{tikzpicture}
\caption{Graph of the non-arrival probability $|r|^2$ as a function of $p$ in units of $\hbar \kappa$.}
\label{reflection probability}
\end{figure}

This arrival time measurement procedure generates a considerable partial quantum Zeno effect for every choice of singular interaction strength $W$ and clock particle velocity $v_c$. To quantify this, consider an incoming plane wave of momentum $p$. Then, the absorbing boundary condition $\partial_x\psi(0)=i\kappa \psi(0)$ produces a reflected plane wave with amplitude 
\begin{equation*}
    \psi=e^{ipx/\hbar}+r(p)e^{-ipx/\hbar} \quad \Rightarrow \quad  r(p)=\frac{p-\hbar \kappa}{p+\hbar \kappa}.
\end{equation*}
It is visible from Figure \ref{reflection probability} that this procedure admits almost no reflection of incoming wavepackets closely centered around $p=\hbar \kappa=4m|W|^2/(v_c\hbar^2)$. The possibility of an exact arrival time measurement procedure with this feature was conjectured earlier by Aharonov et al. \cite{Aharonov}, although they proposed the use of an energy booster that artificially preselects a preferred momentum for detection. Here, $\hbar\kappa$  is an apparatus-dependent quantity that emerges naturally from the particle-detector interaction. 
\par
This partial manifestation of the Zeno effect produces an apparatus-dependent distortion in the arrival distribution which makes this procedure ill-suited to measure certain particle properties, such as the direction of the momentum resulting from a scattering process \cite{Das}. In fact, even the measured arrival time  cannot be viewed as an inherent property of the particle that exists prior to arrival. Instead, as is typical in quantum theory \cite{Bohr1949-BOHDWE,Bell}, the recorded arrival time is a joint product of the particle and its interaction with the apparatus, such that the property it measures can only be associated with the complete experimental setup. This is illustrated by the dependence of the arrival statistics not only on the singular interaction strength $|W|$, but also on the velocity of the clock particle that is used to record the arrival time.
\par
It is notable that the complete quantum Zeno effect can be avoided in this model even as $|W|$ and $v_c$ tend to infinity, so long as the ratio $|W|^2/v_c$ is kept finite and non-zero. The mechanism behind this suppression of the quantum Zeno effect can be understood through a simple finite dimensional model. Consider a two-level system initially prepared in state $|1\rangle $ that undergoes Rabi oscillations with frequency $\Omega$. Suppose also that a third level is introduced to model an apparatus that continuously monitors whether the two-level system has flipped from level $|1\rangle$ to level $|2\rangle$. The simplest model for such an apparatus corresponds to a three-level Hamiltonian of the form
\begin{eqnarray}
    \hat{H}_{3}=\hbar\begin{pmatrix}
        0 & \Omega & 0
        \\
        \Omega & 0 &K
        \\
        0 & K & 0
    \end{pmatrix},
\end{eqnarray}
with $K>0$ representing the registration rate. It is well known that the Zeno effect completely impedes the transition of the two-level system from $|1\rangle$ to $|2\rangle$ as the registration rate $K$ is taken to infinity. However, oscillations can be recovered even as $K \to \infty$ by allowing the apparatus to transition to an additional fourth level $|4\rangle$ after detection \cite{Four, FacchiZeno}, provided the rate of this further transition remains of order $K$. Specifically, the Schr\"odinger dynamics generated by the four-level Hamiltonian
\begin{eqnarray}
    \hat{H}_{4}=\hbar\begin{pmatrix}
        0 & \Omega & 0 &0
        \\
        \Omega & 0 &K&0
        \\
        0 & K & 0 & \alpha K
        \\
        0 & 0 & \alpha K & 0
    \end{pmatrix},
\end{eqnarray}
converges for any fixed $\alpha>0$ as $K$ tends to infinity to $\exp(-it\hat{H}_4/\hbar)|1\rangle \xrightarrow{K \to \infty}\exp(-it\hat{H}_\alpha/\hbar)|1\rangle$ with
\begin{eqnarray}
    \hat{H}_{\alpha}:=\frac{\alpha\hbar}{\alpha^2+1}\begin{pmatrix}
        0 & \alpha\Omega & 0 &-\Omega
        \\
        \alpha\Omega & 0 &0&0
        \\
        0 & 0 & 0 & 0
        \\
        -\Omega & 0 & 0 & 0
    \end{pmatrix}.
\end{eqnarray}
Explicitly, the limiting dynamics can be written as
\begin{eqnarray}
    \exp(-it\hat{H}_{\alpha}/\hbar)|1\rangle =\cos\left(\omega t\right)|1\rangle ~-&i\alpha \mu\sin\left(\omega t\right)|2\rangle\nonumber 
    \\
    +& i\mu\sin\left(\omega t\right)|4\rangle,\nonumber
\end{eqnarray}
where
\begin{equation}
    \omega:=\frac{\alpha\Omega}{\sqrt{1+\alpha^2}}, \quad \mu:=\frac{1}{\sqrt{1+\alpha^2}}\nonumber.
\end{equation}
Oscillations between levels $|1\rangle$ and $|2\rangle$ have been restored in the $K \to \infty$ limit, and a non-zero probability of registration resides in the level $|4\rangle$ sector. 
\par
Detections in this finite dimensional model are not irreversible like the arrival time measurements studied earlier, but this analysis informs why total reflection could be avoided as $|W|$ and $v_c$ tend to infinity. The registration rate relative to the size of the detection region can be arbitrarily large without inducing the full Zeno effect, provided that post-detection states are transferred with sufficient haste after registration to another channel that is not directly coupled to the ready state space.
\par
As a final remark, it is stressed that the production of a reflected wave does not necessarily compromise the precision of an arrival time measurement procedure. It is tempting from classical intuition to associate each reflected wave with an ensemble of free particles whose straight-line trajectories can be traced back to some reflection at $x=0$, but, as Wheeler's delayed choice experiment \cite{Wheeler} shows, such reasoning fails in the presence of interference. There is always interference between the incoming and reflected wave near $x=0$, so whether the detector failed to register the arrival of any reflected particle at $x=0$ is indeterminate according to orthodox quantum theory. Nevertheless, in the models studied here, any non-reflected particle is immediately and irreversibly absorbed upon arrival at $x=0$, and the time of this arrival is exactly recorded in the position of the emitted clock particle. This statement holds even for finite $|W|$ and $v_c$.
\section{Conclusion}
An explicit model for an arbitrarily precise arrival time measurement procedure was presented. The interaction between the incoming particle and the detector was shown to converge to an absorbing boundary condition of the form $\partial_x\psi(0)=4im|W|^2/(v_c\hbar^3)\psi(0)$ as the detection process was localized at $x=0$ and as exact arrival time measurement precision was achieved. Here, $v_c$ denotes the velocity of the clock particle that records the arrival time and $|W|$ controls the detection rate relative to the size of the detection region. A non-zero probability of arrival was retained even as $|W|$ and $v_c$ tend to infinity, provided that the ratio $|W|^2/v_c$ is kept finite and non-zero. This puts to rest any speculation of a fundamental principle in quantum mechanics that denies the possibility of precise ``continuous observation".
\par
A partial quantum Zeno effect always persists in this model that distorts the arrival statistics in an apparatus dependent manner. The dependence of the arrival distribution on both $|W|$ and $v_c$ exemplifies the lesson of Bohr \cite{Bohr1949-BOHDWE} and Bell \cite{Bell} that measuring instruments do not play a passive role in quantum theory. Every component of this arrival time measurement process has become intrinsically linked to the arrival phenomenon.
\par
There are indications that this partial quantum Zeno effect may be reduced with a more sophisticated setup. The model presented here remains amenable to modification and it is possible that the strategic use of a more singular interaction potential, such as those in \cite{Composite,InverseSquare, Complex}, could lead to other arbitrarily precise arrival time measurement procedures with greater absorption.
\begin{acknowledgments}
The author thanks Shadi Tahvildar-Zadeh, Sheldon Goldstein, Roderich Tumulka, Ian Jauslin, Will Cavendish, Rashi Kaimal, and Siddhant Das for many fruitful discussions concerning arrival times in quantum mechanics.
\end{acknowledgments}
\appendix
\section{Convergence of Schr\"odinger dynamics}
Let $\kappa\geq0$. The Laplace transform can be applied to derive an explicit solution to the Schr\"odinger equation in $x<0$ with an absorbing boundary condition at $x=0$
\begin{equation}\label{ABC Schrod}
    i\hbar\partial_t\psi=-\frac{\hbar^2}{2m}\partial_x^2\psi, \quad \partial_x\psi(0)=i\kappa \psi(0).
\end{equation}
Set $\tilde{\psi}_s:=\mathcal{L}\psi=\int_0^\infty dt ~e^{-st}\psi_t$ for $\text{Re}(s)>0$. Taking the Laplace transform of Eq.~(\ref{ABC Schrod}) returns
\begin{equation}
    i\hbar(s\tilde{\psi}_s-\psi_0)=-\frac{\hbar^2}{2m}\partial_x^2\tilde{\psi}_s, \quad \partial_x \tilde{\psi}_s(0)=i\kappa \tilde{\psi}_s(0).
\end{equation}
This equation  is subject to an additional constraint $\lim_{x \to -\infty}\tilde{\psi}_s=0$. An explicit solution formula for $\tilde{\psi}_s$ can be derived and expressed in terms of $\psi_0$ as
\begin{equation}
    \tilde{\psi}_s=G_F+\frac{p-\hbar \kappa}{p+\hbar\kappa}G_R,
\end{equation}
where $p:=\sqrt{2m\hbar i s}$ is defined via the principal square root, so $\text{Re}(ip)<0$, and
\begin{subequations}\label{Greens}
\begin{eqnarray}
    G_F:=&\int_{-\infty}^0 dx'~ \psi_0(x')~\frac{m}{p}e^{ip|x-x'|/\hbar},
\\
    G_R:=&\int_{-\infty}^0 dx'~\psi_0(x')~\frac{m}{p}e^{-ip(x+x')/\hbar}.
\end{eqnarray}
\end{subequations}
The first term corresponds to the free evolution, while the second term contains the reflected contributions. The solution for $\psi_t$ follows from the inverse Laplace transform

\begin{equation}\label{ABC Solution}
    \psi_t=\frac{1}{2\pi \hbar}\int_{-\infty}^{\infty}dE ~e^{-iEt/\hbar}\left(G_F + \frac{p-\hbar\kappa}{p+\hbar\kappa}G_R\right).
\end{equation}
Here, the substitution $E=i\hbar s$ has been performed so that $p(E)=\sqrt{2mE}$.
\par
Now, let $W \in \mathbb{C}$ and $\omega\geq0$. The Laplace transform can also be applied to study the  system of Schr\"odinger equations in $x<0$ coupled through their boundary conditions $x=0$
\begin{subequations}
\begin{equation}
    i\hbar \partial_t\begin{bmatrix}
        \psi\\ \phi
    \end{bmatrix}=\begin{pmatrix}
        -\frac{\hbar^2}{2m}\partial_x^2 &0
        \\
        0&-\frac{\hbar^2}{2M}\partial_x^2-\hbar\omega
    \end{pmatrix}\begin{bmatrix}
        \psi
        \\
        \phi
    \end{bmatrix},
\end{equation}
    \begin{eqnarray}
\partial_x\psi(0)=-\frac{2m}{\hbar^2}W\phi(0), ~ \partial_x\phi(0)=-\frac{2M}{\hbar^2}W^*\psi(0),
    \end{eqnarray}
\end{subequations}
with $\psi\big{|}_{t=0}=\psi_0$ and $\phi \big{|}_{t=0}=0$. Setting $\tilde \psi_s:=\mathcal{L}\psi$ and $\tilde{\phi}_s:=\mathcal{L}\phi$ for $\text{Re}(s)>0$, the transformed equations read
\begin{subequations}
\begin{equation}
    i\hbar \begin{bmatrix}
        s \tilde\psi_s-\psi_0\\ s\tilde\phi_s
    \end{bmatrix}=\begin{pmatrix}
        -\frac{\hbar^2}{2m}\partial_x^2 &0
        \\
        0&-\frac{\hbar^2}{2M}\partial_x^2-\hbar\omega
    \end{pmatrix}\begin{bmatrix}
        \tilde\psi_s
        \\
        \tilde\phi_s
    \end{bmatrix},
\end{equation}
    \begin{eqnarray}
\partial_x\tilde\psi_s(0)=-\frac{2m}{\hbar^2}W\tilde\phi_s(0),~~ \lim_{x \to -\infty}\tilde\psi_s=0, 
\\
\partial_x\tilde\phi_s(0)=-\frac{2M}{\hbar^2}W^*\tilde\psi_s(0),~~ \lim_{x \to -\infty}\tilde{\phi}_s=0.
    \end{eqnarray}
\end{subequations}
Now, the general solution for $\tilde{\phi}_s$ takes the form
\begin{equation*}\tilde\phi_s=Te^{-iqx/\hbar},
\end{equation*}
where $q:=\sqrt{2M\hbar(is+\omega)}$ and $T \in \mathbb{C}$. The boundary condition for $\tilde{\phi}_s$ returns
\begin{equation*}
    T=\frac{-2iM}{\hbar q}W^*\tilde{\psi}_s(0),
\end{equation*}
which reduces the boundary value problem for $\tilde{\psi}_s$ to
\begin{subequations}
    \begin{equation}
        i\hbar (s\tilde\psi_s - \psi_0)=-\frac{\hbar^2}{2m}\partial_x^2\tilde{\psi}_s
    \end{equation}
    \begin{equation}
        \partial_x\tilde{\psi}_s(0)=i\frac{4mM}{\hbar^3q}|W|^2\tilde{\psi}_s(0), \quad \lim_{x \to -\infty}\tilde\psi_s=0.
    \end{equation}
\end{subequations}
Introducing $\alpha:=4mM|W|^2/(\hbar^3 q)$, the solution formula for $\tilde{\psi}_s$ can now be expressed as
\begin{equation}
    \tilde{\psi}_s=G_F+\frac{p-\hbar\alpha}{p+\hbar\alpha}G_R,
\end{equation}
where $p=\sqrt{2m\hbar i s}$ and $G_F$ and $G_R$ are defined in Eq.~(\ref{Greens}). The inverse Laplace transform returns
\begin{equation}\label{dynamics}
    \psi_t=\frac{1}{2\pi \hbar}\int_{-\infty}^{\infty}dE ~e^{-iEt/\hbar}\left(G_F+\frac{p-\hbar\alpha}{p+\hbar\alpha}G_R \right).
\end{equation}
Here, $p(E)=\sqrt{2mE}$ and $q(E)=\sqrt{2M(E+\hbar \omega)}$. Now, let $v_c>0$ and set $M=2\hbar\omega/v_c^2$. Then
\begin{equation*}
    q/M=\sqrt{2(E+\hbar\omega)/M}=\sqrt{v_c^2 +(Ev_c^2/\hbar\omega}).
\end{equation*}By dominated convergence theorem, the limit as $\omega \to \infty$ of the solution formula (\ref{dynamics})  converges to Eq.~(\ref{ABC Solution}) with $\kappa=4m|W|^2/(\hbar^3 v_c)$, as desired. 
\par
To analyze the dynamics of the post-detection state $\phi_t$, one may employ the stationary phase method. Recall that $\tilde{\phi}_s$ was given by
\begin{equation}
    \tilde{\phi}_s= \left(\frac{M}{q}e^{-iqx/\hbar}\right) \left( \frac{-2i}{\hbar}W^*\tilde{\psi}_s(0)\right),
\end{equation}
where $-iq/\hbar=-i\sqrt{2M(is+\omega)/\hbar}=\sqrt{-2Mi(s-i\omega)/\hbar}$ since $\text{Re}(s)>0$. Then, setting $M=2\hbar\omega/v_c^2$, the post-detection state can be expressed as a convolution
\begin{equation}
    \phi_t=-\frac{2W^*}{\hbar v_c}\sqrt{\frac{i\omega}{\pi}}\int_{0}^t d\tau ~\frac{\psi_{t-\tau}(0)}{\sqrt{\tau}}~e^{i\omega( \tau + x^2/(v_c^2\tau))}.
\end{equation}
For $\omega$ sufficiently large, this can be approximated using the stationary phase method as
\begin{equation}
    \phi_t\approx\begin{cases}
        \frac{-2iW^*}{\hbar v_c}e^{-2i\omega x/v_c}\psi_{t+x/v_c}(0) \quad &x>-v_ct
        \\
        0 \quad &x<-v_ct
    \end{cases}
\end{equation}
plus terms of order $O(\omega^{-1})$.
Hence, the limiting dynamics for $\psi$ and $\phi$ is governed by Eq~(\ref{stopwatch})
as desired.

\bibliography{ToA}
\end{document}